\documentclass[11pt]{article}
\usepackage{moriond,epsfig}

\def\be{\begin{equation}}
\def\ee{\end{equation}}
\def\bea{\begin{eqnarray}}
\def\eea{\end{eqnarray}}

\begin{document}
\vspace*{4cm}
\title{LATEST JET RESULTS FROM THE TEVATRON}

\author{ DARREN D. PRICE \\
on behalf of the CDF and D\O\ Collaborations}

\address{Department of Physics, Indiana University,\\
Bloomington, IN 47405, USA}

\maketitle\abstracts{
A brief overview of the latest status of jet physics studies at the Tevatron in proton-antiproton collisions at $\sqrt{s}=1.96$~TeV is presented.
In particular, measurements of the inclusive jet production cross-section, dijet production and searches for new physics,
the ratio of the 3-jet to 2-jet production cross-sections, and the three-jet mass are discussed.
}

The measurement of inclusive jet rates in Tevatron data at $\sqrt{s}=1.96$~TeV has allowed for tests of perturbative QCD (pQCD) and searches
for new physics at jet transverse momenta of up to 700~GeV and over eight orders of magnitude in cross-section. 
Much work went into understanding the calorimeter response in both experiments,
using a single particle response technique\,\cite{CDF-JES} in the case of CDF and a data-driven photon-plus-jet event calibration
method at D\O. Both extended the calorimeter response to non-optimal calorimeter regions using dijet $p_T$ balancing techniques. 
The result of such studies and accurate simulation was to minimise the experimental systematics and these inclusive measurements are thus dominated 
by theoretical uncertainties. As a result, the Tevatron has been able to make significant contributions to the understanding of the proton structure 
and the strong force and improve sensitivity to new physics effects. The understanding gained by these measurements are important not just for QCD 
analyses, but also have relevance to any analyses which have jets as a feature of interest.

CDF's measurement of the inclusive jet cross-section was performed\,\cite{CDF-midpoint} using the midpoint cone algorithm\,\cite{midpoint} 
with a cone size of $R=0.7$ in five bins of jet rapidity up to $|y|<2.1$ and, as in all studies described here, was fully corrected for efficiencies and 
bin-to-bin migrations caused by the $p_T$ resolution of the detector. The corrected spectrum was compared to NLO pQCD from
\textsc{Fast}NLO\,\cite{FastNLO} based on the NLOJet++\,\cite{NLOJet} program, using CTEQ6.1M\,\cite{CTEQ61M}. 
NLO calculations are provided
at the parton-level whilst cross-sections are corrected back to the particle-level. As such, in all comparisons to NLO a parton-to-particle non-perturbative 
correction is derived from parton-shower Monte Carlo (and applied to the NLO prediction) to place the corrected data and the NLO theory on an equal footing.
Such corrections are largest
at low jet $p_T$, where underlying event corrections to the jet area can be significant (of order $10-20$\%). After this correction,
the measured cross-sections were found to be lower than but in agreement with NLO within the uncertainties.
CDF also measured\,\cite{CDF-kt} the inclusive jet cross-section using the $k_T$ clustering algorithm\,\cite{kt} for three jet size parameter choices
$D=0.4$, $0.7$ and $1.0$. An advantage of this jet algorithm is its infrared and collinear safety to all orders in perturbation theory, and measurement
using both $k_T$ and midpoint is an important validation test of the use of different jet algorithms at hadron colliders. The data with $k_T$ and 
midpoint were found to agree across a wide range of rapidity and $p_T$. NLO theory and data were also in good agreement 
apart from in the highest rapidity bin where the data is lower than NLO prediction (but within uncertainties),
with the measurements using different distance parameters showing similar behaviours. From this one may conclude that the cone
and $k_T$ clustering algorithms can be used to retrieve consistent results at hadron-hadron machines.

D\O's measurement\,\cite{D0-midpoint} of the inclusive jet cross-section made use of the midpoint cone algorithm with $R=0.7$, analysed data in 
six rapidity bins up to $|y|<2.4$ and is the most precise measurement to-date. A comparison was also made to NLO theory with NLOJet++ and 
\textsc{Fast}NLO using the CTEQ6.5M PDFs, and whilst in agreement, the data prefers the lower bound of the theoretical prediction. 
The jet energy scale (JES) uncertainty of $1.2-2$\% (compared to CDF's $2-3$\%) dominates the experimental error. 
Due to the steeply falling cross-section this translates into a large error on the final results, leading to total errors 
on the measurement of $15-30$\% for D\O\ and $15-50$\% for CDF. These uncertainties are generally 
smaller than those from theory (largely coming from PDF uncertainties), and has enabled (along with the CDF inclusive jet data) 
constraints of the gluon PDF at high $x$ and high $Q^2$, now used in the MSTW2008\,\cite{MSTW2008} fits.

\begin{figure}[htb]
  \begin{center}
   \includegraphics[width=0.41\textwidth, trim= 0cm 0cm 0cm 0.1cm, clip]{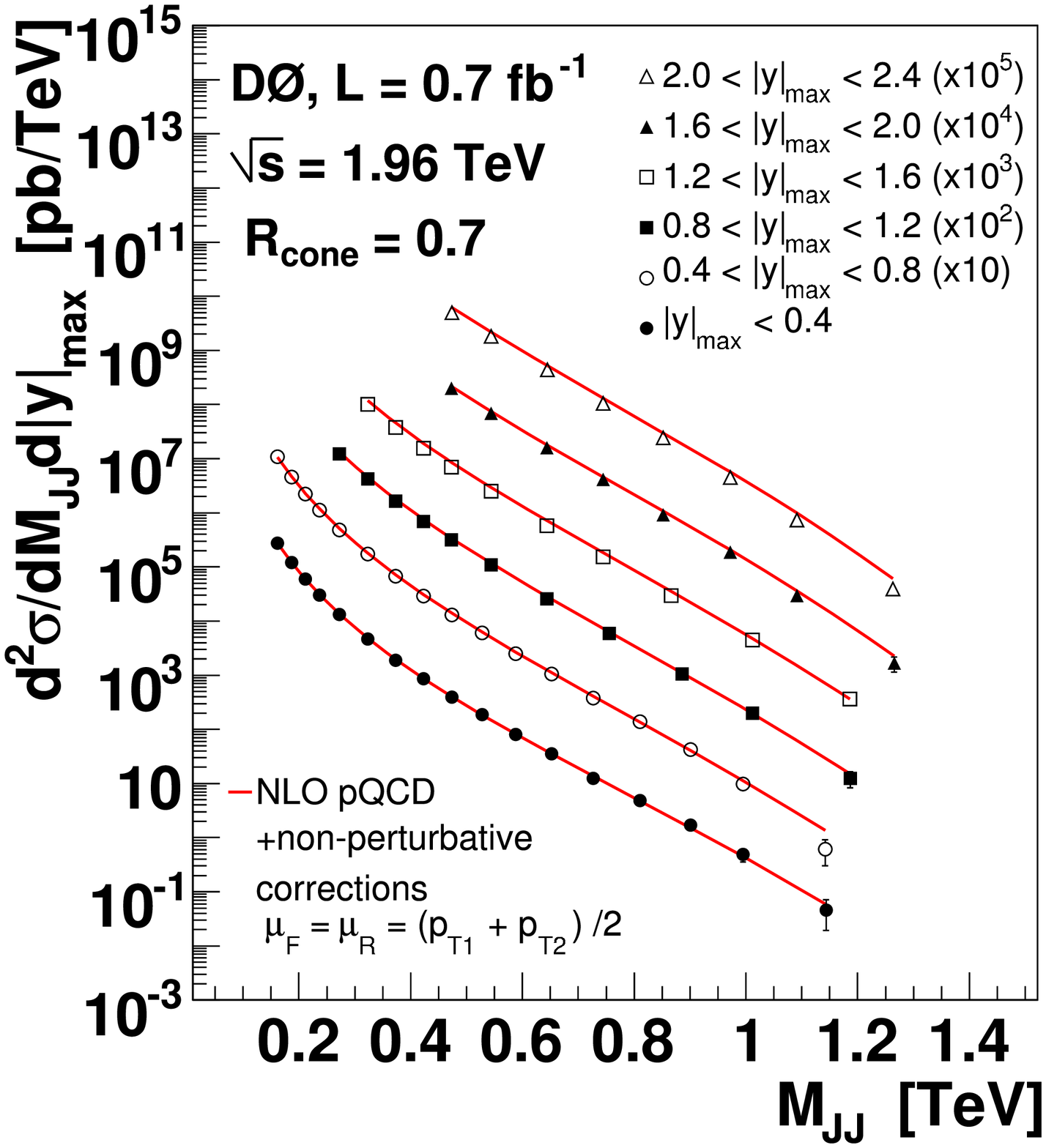}
   \includegraphics[width=0.36\textwidth, trim= 0cm 2cm 0cm 2cm, clip]{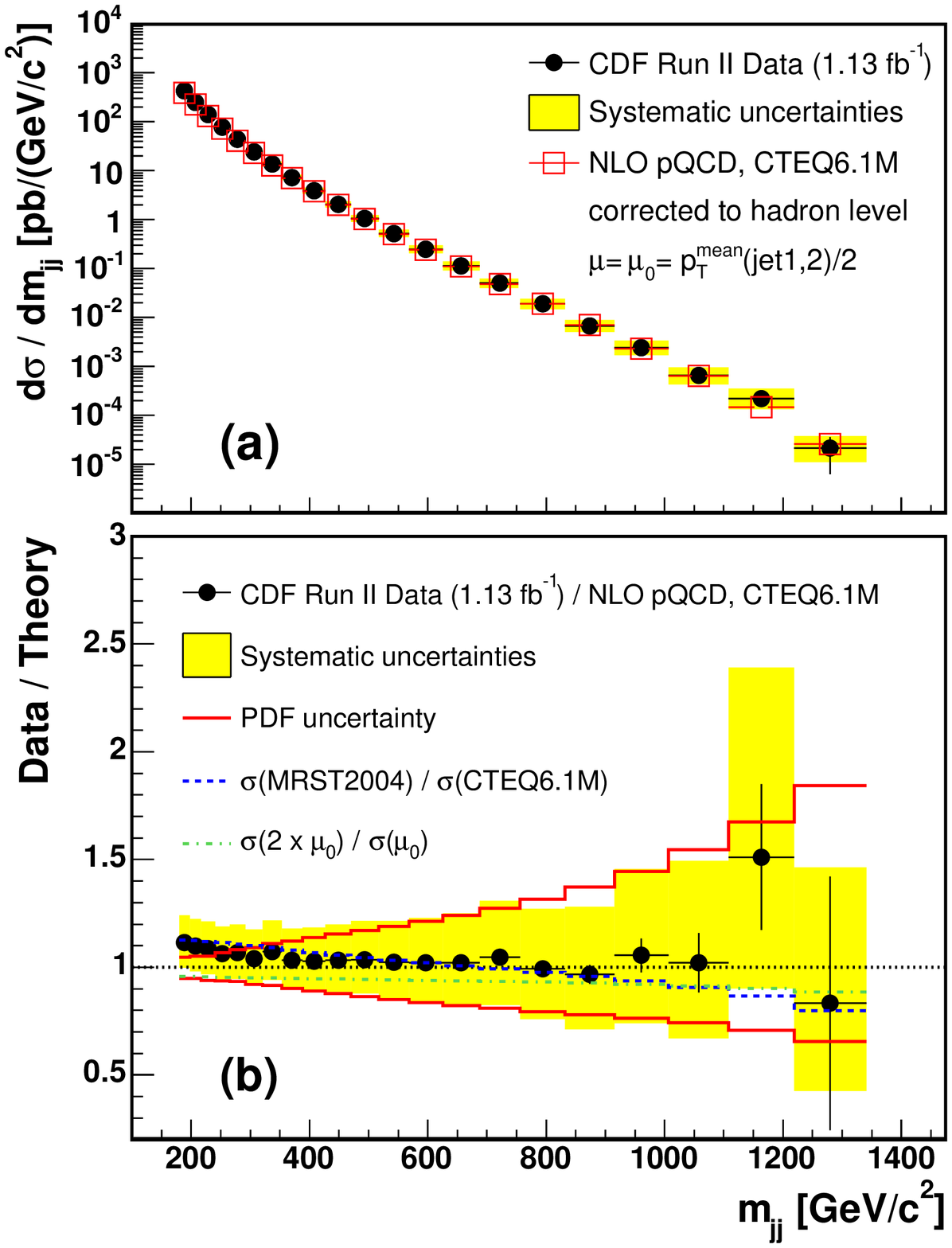}
   \caption{Measured dijet mass cross-section from D\O\ (in bins of the highest rapidity jet) {\em(left)} and from CDF for jets in $|y|<1.0$
     {\em(right)} compared to NLO calculations. In both cases the theoretical uncertainty from the PDF is 
     comparable to the systematic uncertainties (largely from the jet energy scale uncertainty).
   }
   \label{fig:dijet}
  \end{center}
\end{figure}

\begin{figure}[htb]
  \begin{center}
    \includegraphics[width=0.33\textwidth, trim= 0cm 5.5cm 0cm 3cm, clip]{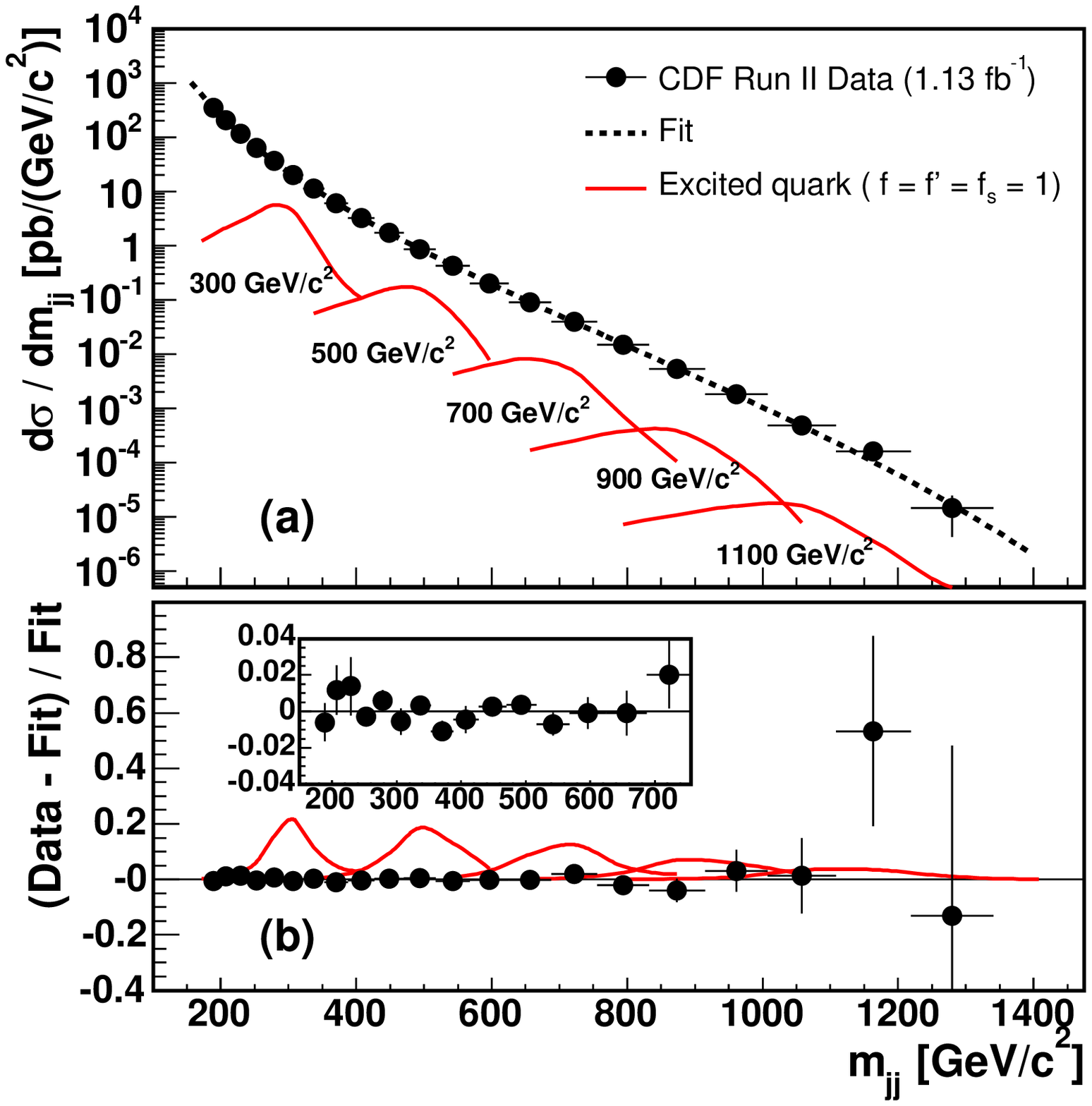}
    \includegraphics[width=0.48\textwidth, trim = 0cm 0cm 0cm 0cm, clip]{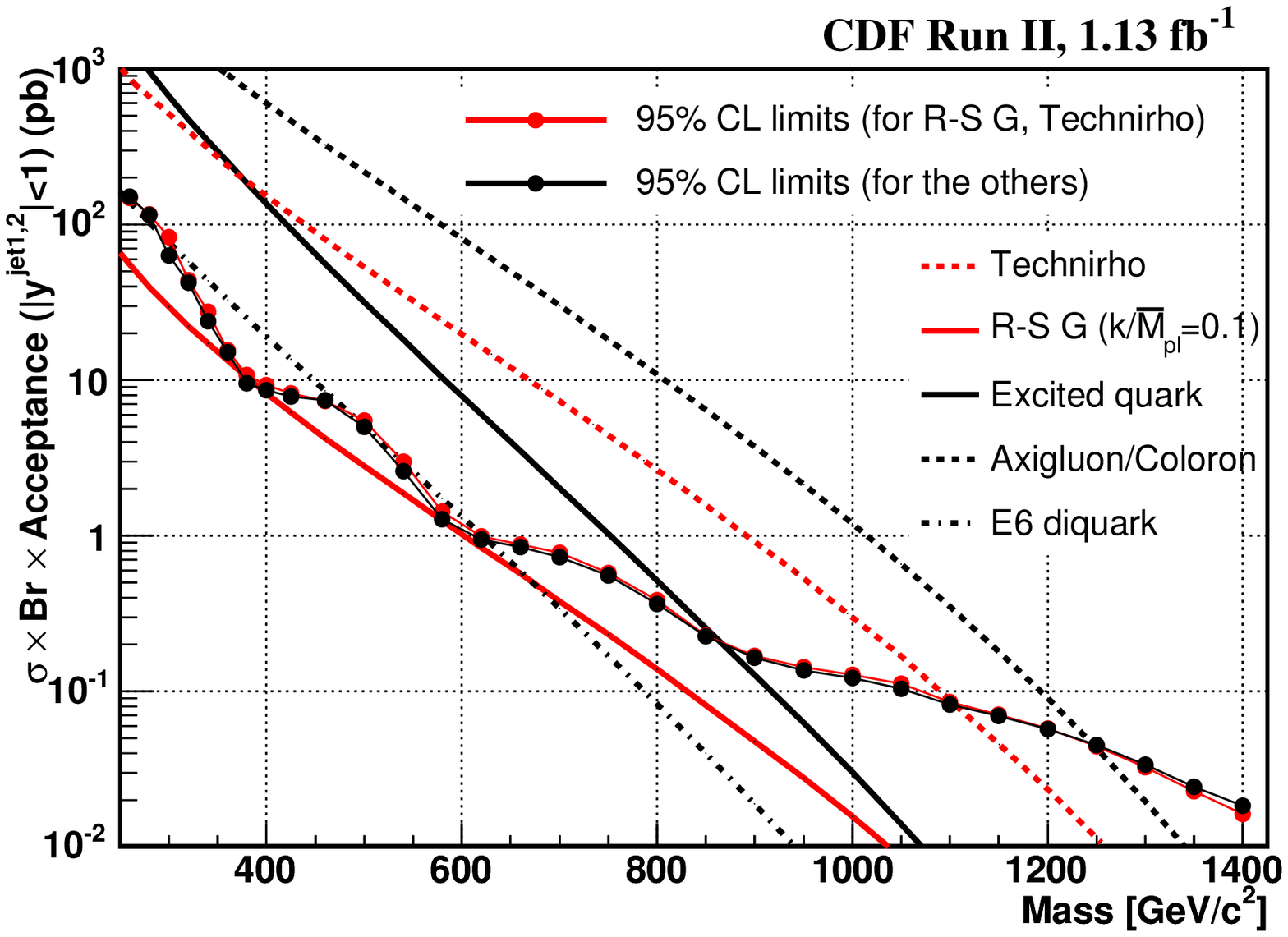}
    \caption{Measured dijet mass spectrum before bin-by-bin unfolding (a) and ratio (b) compared to excited quark signals 
      expected at various masses {\em (left)}, and resultant 95\% C.L limits for various BSM predictions {\em (right)}.
    }
    \label{fig:mjjBSM}
  \end{center}
\end{figure}

Utilising this well-understood dataset it is possible to extract many other jet results. Both CDF\,\cite{CDF-dijet} and D\O\,\cite{D0-dijet}
measured the dijet mass spectrum (see Figure~\ref{fig:dijet}), not only as a test of theoretical calculations but as a search for new physics in
models that predict the existence of a particle that decays into two high $p_T$ jets. 
D\O\ made a measurement of the dijet mass in the six 
rapidity bins of the inclusive cross-section measurement and compared the results to NLO predictions from \textsc{Fast}NLO with 
MSTW2008NLO PDFs. Bin sizes in $m_{JJ}$ are chosen to give a bin purity and efficiency\footnote{Determined from a parameterised detector model. 
Efficiency [purity] is defined as the ratio of Monte Carlo events reconstructed [generated] to those generated [reconstructed] in a particular bin.} 
of about 50\%; experimental corrections vary between 0.5\% and 12\%, NLO non-perturbative corrections are between $5-20$\% in size. 
Systematic uncertainties on the measurement are similar to those from PDF and scale uncertainties, suggesting the measurement 
can be used to constrain future predictions. 
CDF restricted itself to a central jet ($|y|<1.0$) measurement (also shown in Figure~\ref{fig:dijet})
where jets from new physics are most likely to be produced. The data were consistent
with NLO predictions. From this data, CDF searched for narrow dijet resonances by fitting the data before bin-by-bin unfolding corrections
(to avoid any resultant degradation in a possible signal) to a smooth functional form and looking for significant data excesses from the fit. 
Figure~\ref{fig:mjjBSM} shows the expected signals in the presence of excited quarks at various masses, and in the absence of any resonant
structure, exclusion limits for various new physics models and in particular the most stringent limits on excited quark, axigluon,
flavour-universal coloron, $E_6$ diquark and colour-octet techni-$\rho$ models.

\begin{figure}[htb]
  \begin{center}
    \includegraphics[width=0.4\textwidth, trim = 1.5cm 0cm 1.0cm 1cm, clip]{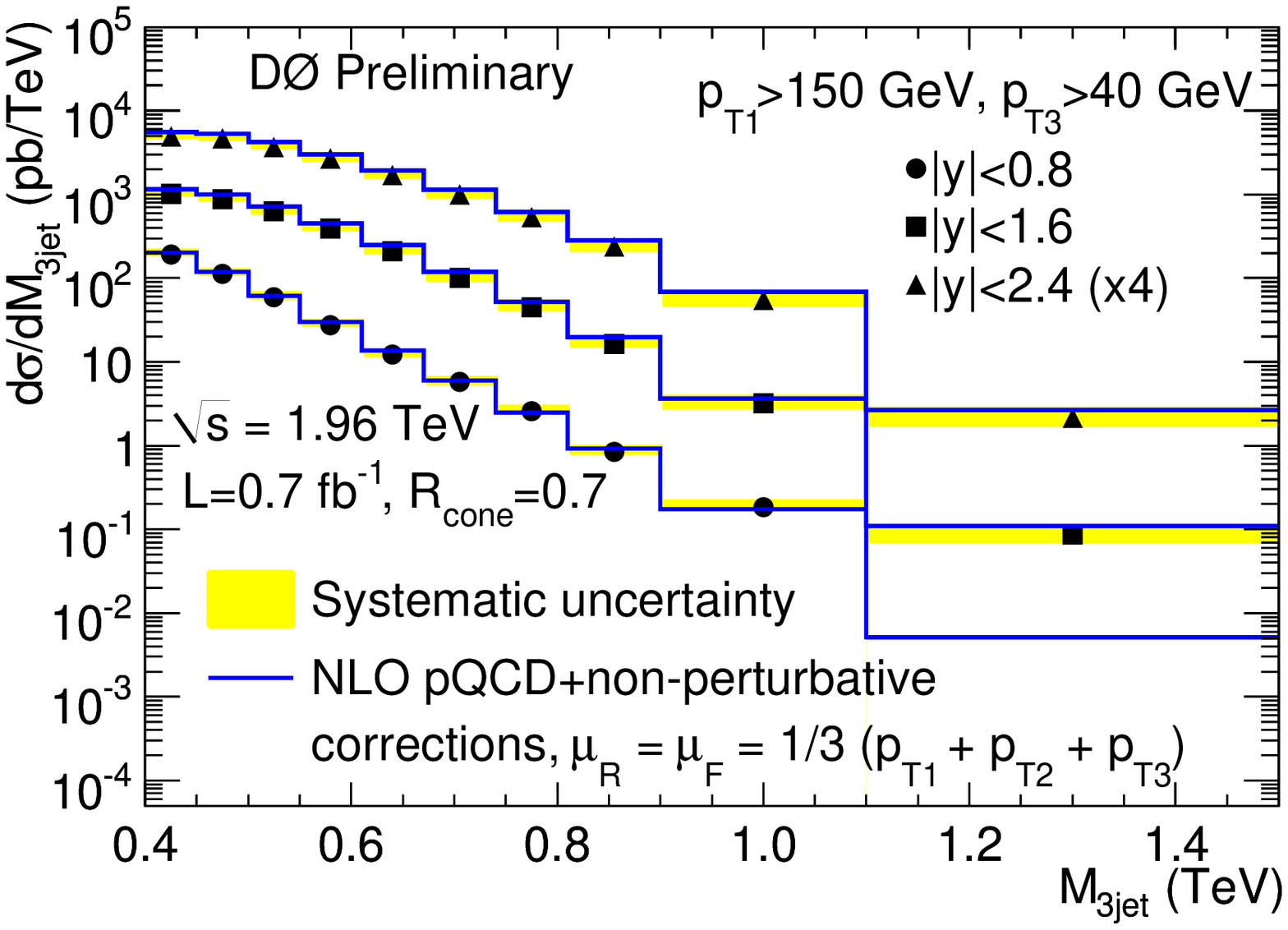}
    \includegraphics[width=0.4\textwidth, trim = 1.0cm 0cm 1.5cm 1cm, clip]{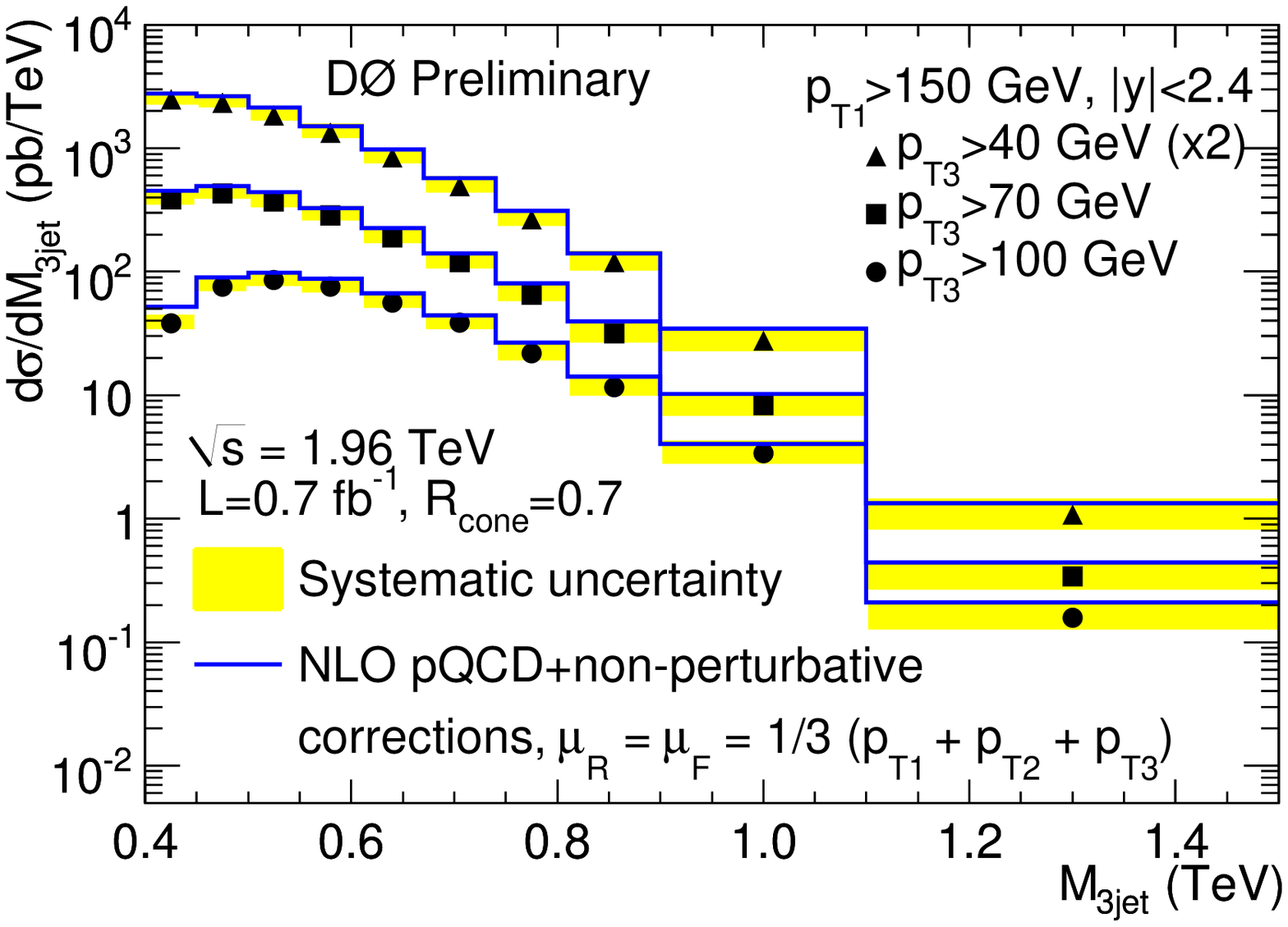}
    \caption{Three-jet mass cross-section in bins of jet rapidity {\em (left)} and third jet $p_T$ {\em (right)}
      and systematic uncertainties (up to $20-30$\%) compared to NLO calculations using NLOJet++ and MSTW2008LO PDFs.
    }
    \label{fig:jjj}
  \end{center}
\end{figure}

D\O\ has made the first measurement of the three-jet cross-section in RunII data from the Tevatron (see also a previous result\,\cite{CDF3jet}
from CDF using RunI data) in three rapidity regions ($|y|<0.8, <1.6, <2.4$) 
and three bins ($p_{T3}>40, 70, 100$~GeV) of third jet transverse momentum (shown in Figure~\ref{fig:jjj}) as a function of three-jet mass. 
A leading jet requirement of $p_{T1}>150$~GeV, 
in conjunction with the third jet $p_T$ requirement ensures the trigger for selected events is 100\% efficient. Any pair $ij$
of jets is required to have a $y-\phi$ spatial separation of $\Delta R_{ij}>1.4$ to avoid jet overlap reliant on the split-merge
procedure of the midpoint jet cone algorithm. The \textsc{Sherpa}\,\cite{Sherpa} Monte Carlo generator with MSTW2008LO PDFs were used to perform the
event simulation and correct for various detector effects after determining the generator performed well at describing the data distributions.
There is agreement with NLO calculations, with the data preferring the lower range of the predictions.

\begin{figure}[htb]
  \begin{center}
    \includegraphics[width=0.70\textwidth, trim = 0cm 1cm 0cm 0cm, clip]{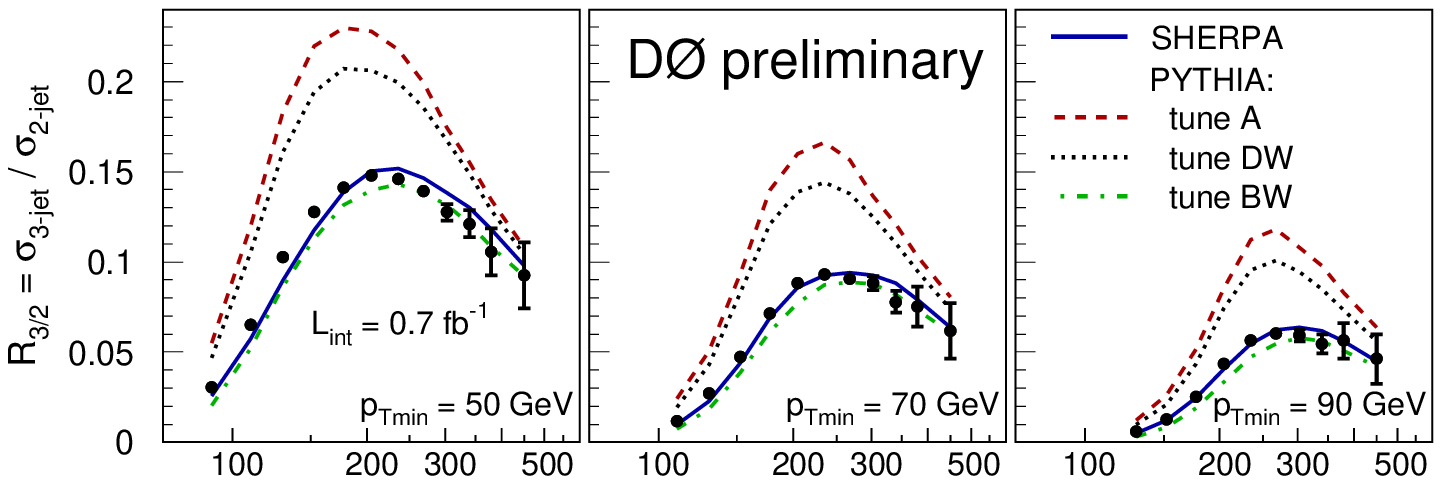}
    \includegraphics[width=0.70\textwidth, trim = 0cm 0.1cm 0cm 0cm, clip]{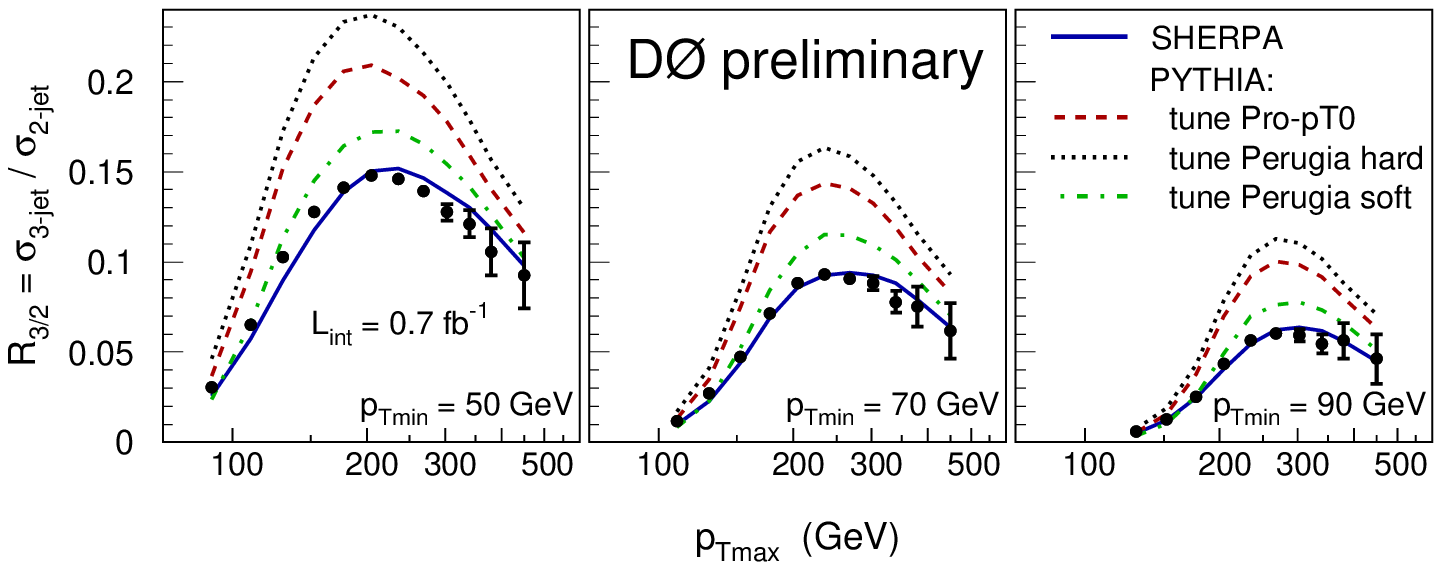}
    \caption{Ratio of trijet to dijet cross-section with hardest jet $p_T$ in bins of $p_{T\textrm{min}}$ of
      the other jets and predictions of \textsc{Sherpa} and of \textsc{Pythia} for three tunes with virtuality-ordered
      showers {\em (top)} and $p_T$-ordered showers {\em (bottom)}.
    }
    \label{fig:r32}
  \end{center}
\end{figure}

Using the same dataset at D\O\ and again using simulation with \textsc{Sherpa}, the first measurement of the ratio of the three to 
two-jet cross-section at the Tevatron has been made corrected for all detector effects and 
measured as a function of two momentum scales: $p_{T\textrm{max}}$ the leading jet $p_T$
and $p_{T\textrm{min}}$ the scale at which the other jets are resolved. This is a test of pQCD largely independent of PDFs, and many other 
uncertainties cancel in the ratio making this measurement particularly sensitive. The results are shown in Figure~\ref{fig:r32} over a range
$p_{T\textrm{min}}+30$~GeV$<p_{T\textrm{max}}<500$~GeV to allow sufficient phase space for jets to be resolved and experimental corrections to
be small ($0.9-1.2$ in the ratio). A jet $\Delta R_{ij}>1.4$ requirement is again used to ensure good separation of the jets. Despite the
relatively small integrated luminosity, the measurement is dominated by systematic uncertainties ($<5$\%) for $p_{T\textrm{max}}<250-300$~GeV.

This ratio can be interpreted as the conditional probability for an inclusive dijet event at $p_{T\textrm{max}}$ to contain a third jet.
\textsc{Sherpa} predictions using MSTW2008LO (with matrix element matching for up to 4-jet production)
are compatible with data within 20\%, but \textsc{Pythia} comparisons (which rely on the parton
shower for more than two jet emissions) are generally unable to describe the data. Tune BW has a reasonable description,
but not as good as that of \textsc{Sherpa}, and is incompatible with D\O\ measurement of dijet azimuthal decorrelations.

\end{document}